\documentclass[10pt,prd,twocolumn,
nofootinbib,
noshowkeys,
noshowpacs,
superscriptaddress,
floatfix
]{revtex4-2}
\usepackage{amsmath,amsfonts,amsthm,amssymb}
\usepackage[dvips]{graphics,graphicx}
\usepackage[usenames,dvipsnames]{color}
\definecolor{darkblue}{RGB}{0,0,196}
\definecolor{darkgreen}{RGB}{0,120,0}
\usepackage[colorlinks=true,linktocpage=true,linkcolor=darkblue,citecolor=red,urlcolor=darkblue]{hyperref}
\usepackage{cancel}
\usepackage{bbold}
\usepackage{multirow}
\usepackage{longtable}
\usepackage{color}
\usepackage[normalem]{ulem}
\usepackage{hyperref}
\usepackage{bigints}
\usepackage{xparse}
\usepackage{physics}
\usepackage{verbatim}
\usepackage{minibox}
\usepackage{comment}
\usepackage{appendix}
\usepackage{slashed}
\usepackage{marginnote}
\usepackage{graphicx}
\usepackage[nice]{nicefrac}
\usepackage{amsmath}
\usepackage{hepunits}
\include{commands}
\usepackage{stackengine,scalerel}
\newcommand\hstar[1]{\ThisStyle{\ensurestackMath{%
  \setbox0=\hbox{$\SavedStyle#1$}%
  \stackengine{0pt}{\copy0}{\kern.2\ht0\smash{\SavedStyle\star}}{O}{c}{F}{T}{S}}}}

\definecolor {darkgreen}{rgb}{0.2,0.7,0.2}

\begin{document}
\title{Boost invariant spin hydrodynamics within the first order in derivative expansion}

\author{Rajesh Biswas}
\email{rajeshbiswas@niser.ac.in}
\affiliation{School of Physical Sciences, National Institute of Science Education and Research, An OCC of Homi Bhabha Nuclear Institute, Jatni-752050, India}

\author{Asaad Daher}
\email{asaad.daher@ifj.edu.pl}
\affiliation{Institute  of  Nuclear  Physics  Polish  Academy  of  Sciences,  PL-31-342  Krak\'ow,  Poland}

\author{Arpan Das}
\email{arpan.das@ifj.edu.pl}
\affiliation{Institute  of  Nuclear  Physics  Polish  Academy  of  Sciences,  PL-31-342  Krak\'ow,  Poland}

\author{Wojciech Florkowski}
\email{wojciech.florkowski@uj.edu.pl}
\affiliation{Institute of Theoretical Physics, Jagiellonian University, PL-30-348 Krak\'ow, Poland}

\author{Radoslaw Ryblewski}
\email{radoslaw.ryblewski@ifj.edu.pl}
\affiliation{Institute  of  Nuclear  Physics  Polish  Academy  of  Sciences,  PL-31-342  Krak\'ow,  Poland}







\begin{abstract}
	Boost-invariant equations of spin hydrodynamics confined to the first-order terms in gradients are numerically solved. The spin equation of state, relating the spin density tensor to the spin chemical potential, is consistently included in the first order. Depending on its form and the structure of the spin transport coefficients, we find solutions which are both stable and unstable within the considered evolution times of 10 fm/c. These findings are complementary to the recent identification of stable and unstable modes for perturbed uniform spin systems described by similar hydrodynamic frameworks.

\end{abstract}

\date{\today \hspace{0.2truecm}}

 \maketitle
\flushbottom







%
\section{Introduction}
Recent evidence of the spin polarization of weakly decaying Lambda hyperons has opened up a new pathway for investigating non-trivial vortical structures of strongly interacting matter generated in heavy-ion experiments~\cite{STAR:2017ckg, STAR:2018gyt,STAR:2019erd,ALICE:2019onw,ALICE:2019aid,STAR:2020xbm,Kornas:2020qzi,STAR:2021beb,ALICE:2021pzu,lisa2021}. Several approaches for describing relativistic hydrodynamics for spin-polarized fluids have been developed as a result of the successes of the relativistic dissipative hydrodynamic framework in heavy-ion phenomenological research~\cite{Florkowski:2017olj,Elfner:2022iae}.~Different frameworks have been constructed using entropy current analysis~\cite{Hattori:2019lfp,Fukushima:2020ucl,Li:2020eon,Hongo:2021ona,She:2021lhe,Wang:2021ngp,Daher:2022xon}, relativistic kinetic theory~\cite{Florkowski:2017ruc,Florkowski:2017dyn,Florkowski:2018ahw,Florkowski:2018fap,Florkowski:2019qdp,Bhadury:2020puc,Bhadury:2022qxd,Fang:2022ttm,Bhadury:2020cop,Speranza:2020ilk,Weickgenannt:2020aaf,Weickgenannt:2021cuo,Shi:2020htn,Peng:2021ago,Sheng:2021kfc,Weickgenannt:2022qvh}, effective Lagrangian approach~\cite{Montenegro:2017rbu,Montenegro:2017lvf,Montenegro:2018bcf,Montenegro:2020paq}, quantum statistical density operators~\cite{Becattini:2007nd,Becattini:2009wh,Becattini:2012pp,Becattini:2018duy,Hu:2021lnx}, equilibrium partition functions~\cite{Gallegos:2021bzp}, and holography~\cite{Gallegos:2020otk,Garbiso:2020puw}. To allow for future dynamic simulations of spin polarization~\cite{Becattini:2022zvf, Wagner:2022gza,Hu:2022xjn}, a consistent framework of relativistic hydrodynamics with spin degrees of freedom (spin hydrodynamics) is currently being built.

In this work, we present an analysis of boost-invariant solutions of the spin hydrodynamic equations formulated by Hattori et al. in~Ref.~\cite{Hattori:2019lfp} and investigated later in a series of publications~\cite{Fukushima:2020ucl,Wang:2021ngp,Daher:2022xon,Sarwar:2022yzs,Daher:2022wzf}. The approach of Ref.~\cite{Hattori:2019lfp} is based on the gradient expansion and requirement of positive entropy production. This leads to the identification of first-order in gradients corrections to the energy-momentum tensor $T^{\mu\nu}$, in a way similar to the construction of the Navier-Stokes relativistic hydrodynamics for ordinary (spinless) fluids. A novel feature of spin hydrodynamics is that the gradient corrections include not only symmetric but also antisymmetric contributions to $T^{\mu\nu}$. The presence of such antisymmetric parts of the energy-momentum tensor leads to a non-trivial equation for the spin tensor of the fluid, which should be treated as one of the hydrodynamic equations (in addition to the standard conservation of $T^{\mu\nu}$). 

A characteristic feature of the framework proposed in Ref.~\cite{Hattori:2019lfp} is also the form of the leading (zeroth order) contribution to the spin tensor,
\begin{equation}
S^{\mu\alpha\beta}_{\rm ph}=u^{\mu} S^{\alpha \beta},
\label{eq:spintenph}
\end{equation}
which can be traced back to the seminal work of Weyssenhoff and Raabe~\cite{Weyssenhoff:1947iua} (here $u^\mu$ is the hydrodynamic flow vector and the antisymmetric tensor $S^{\alpha \beta}$ describes spin density). The formulation of spin hydrodynamics with such a form of the spin tensor is often called the phenomenological approach. Recently, a connection between this form and the canonical formalism of spin hydrodynamics, where the spin tensor is totally antisymmetric, has been established~\cite{Daher:2022xon}. It turns out that these two approaches differ not only by a pseudo-gauge transformation~\cite{HEHL197655} but also by a gradient term which should be included in the definition of the canonical energy-momentum tensor. Nevertheless, the structure of those differences makes the phenomenological and canonical formulations of spin hydrodynamics completely equivalent. The results of Ref.~\cite{Daher:2022xon} shed new light also on the results presented in Ref.~\cite{Fukushima:2020ucl} where the Belinfante form of the energy-momentum tensor is used --- as long as the derivation of the dissipative spin hydrodynamics starts from the same definition of the entropy current, the resulting hydrodynamic framework is the same as in Ref.~\cite{Hattori:2019lfp}. Given these findings, in this work, we continue to work with the form~(\ref{eq:spintenph}). 

Another general problem one encounters while dealing with the gradient expansion for spin hydrodynamics is the counting scheme --- the same physical quantities may be considered to be of different order (in the gradient expansion) in different works. Herein, we follow the strategy outlined in~\cite{Hattori:2019lfp} and treat the spin density $S^{\alpha\beta}$ as being of the zeroth order in gradients $(\mathcal{O}(1))$, and the spin chemical potential $\omega^{\alpha\beta}$ to be of the first order $(\mathcal{O}(\partial))$. This leads, however, to a substantial difficulty if one assumes that $S^{\alpha\beta}$ is a function of temperature $T$ and $\omega^{\alpha\beta}$, and uses the equation $S^{\alpha\beta}(T,\omega^{\alpha\beta}) = S_0(T) \omega^{\alpha\beta}$, as the orders of magnitude of the two sides of this equation do not match ($S_0(T)$ here is a certain function of temperature). A possible solution to this problem is to argue that one considers the case where $S_0(T)$ is so large that it compensates for the smallness of $\omega^{\alpha\beta}$~\cite{Hattori:2019lfp}. This leads us, however, beyond the original gradient expansion. In this work, to remain within the gradient expansion we assume the following dependence 
\begin{align}
    S^{\mu\nu}(T,\omega^{\mu\nu})=S_{0}(T)\frac{\omega^{\mu\nu}}{\sqrt{\omega^{\mu\nu}\omega_{\mu\nu}}}.
    \label{spintensor0}
\end{align}
A more detailed discussion of the form (\ref{spintensor0}) will be given below, here we only stress that Eq.~(\ref{spintensor0}) is a new feature explored in this work that makes it different from previous studies performed within the first order spin hydrodynamics. In particular, this makes our analysis different and complementary to an earlier work on boost-invariant solutions of spin hydrodynamics performed in Ref.~\cite{Wang:2021ngp}. The formalism developed herein is also similar to that presented in Refs.~\cite{Florkowski:2019qdp,Florkowski:2021wvk} where, however, a significantly different formulation of spin hydrodynamics was used. 

\medskip
Our paper is organized as follows: In Sec.~\ref{sec:spinhydro} we define the framework of spin hydrodynamics that is based on the gradient expansion, introduce the constitutive equations for matter with spin polarization, and identify the form of the dissipative currents. In Sec.~\ref{sec:implementBI} the symmetry of boost-invariance is implemented into our hydrodynamic framework. The forms of the spin equation of state and the spin kinetic coefficients are presented in Sec.~\ref{sec:EOS}. The results of our numerical calculations are given in Sec.~\ref{sec:numerics}. Finally, we conclude in Sec.~\ref{sec:conclusions}.

\medskip
In this work we use the notation where $\epsilon^{\mu\nu\alpha\beta}$ is the totally antisymmetric tensor and we follow the convention $\epsilon^{0123}=-\epsilon_{0123}=1$.

\section{Spin hydrodynamics}
\label{sec:spinhydro}
\subsection{Basic conservation laws}

The hydrodynamic framework for a spin-polarized fluid is based on the 
conservation laws for the energy-momentum tensor $T^{\mu\nu}$ and the total angular momentum tensor $J^{\lambda\mu\nu}$,
\begin{align}
    \partial_{\mu}T^{\mu\nu}=0,\\
    \partial_{\lambda}J^{\lambda\mu\nu}=0.
\end{align}
The total angular momentum tensor is the sum of the orbital part $L^{\lambda\mu\nu}=2\,x^{[\mu}T^{\lambda\nu]}$, and the spin part  $S^{\lambda\mu\nu}$\footnote{Symmetric and antisymmetric part of a tensor $X^{\mu\nu}$ is denoted as 
 $X^{\mu\nu}_{(s)}\equiv X^{(\mu\nu)}=(X^{\mu\nu}+X^{\nu\mu})/2$ and $X^{\mu\nu}_{(a)}\equiv X^{[\mu\nu]}=(X^{\mu\nu}-X^{\nu\mu})/2$, respectively.
}. We use the phenomenological energy-momentum and spin tensors which have the following forms~\cite{Hattori:2019lfp}
\begin{align}
        & T^{\mu\nu}_{\rm ph}=\varepsilon u^{\mu}u^{\nu}-p\Delta^{\mu\nu}+h^{\mu}u^{\nu}+h^{\nu}u^{\mu}\nonumber\\
        & ~~~~~~~~~+\tau^{\mu\nu}+q^{\mu}u^{\nu}-q^{\nu}u^{\mu}+\phi^{\mu\nu}, \label{Tmn} \\
        & S^{\mu\alpha\beta}_{\rm ph}=u^{\mu} S^{\alpha \beta}+S^{\mu\alpha\beta}_{(1)}. \label{Slmn}
\end{align}
Here $\varepsilon$ is the energy density, and $p$ is the equilibrium pressure. We define $u^{\mu}$ as the fluid four-velocity satisfying the normalization condition $u^{\mu}u_{\mu}= 1$, and $\Delta^{\mu\nu}$ is the symmetric operator projecting onto the space orthogonal to $u^{\mu}$, i.e., $\Delta^{\mu\nu}=g^{\mu\nu}-u^{\mu}u^{\nu}$, where $g_{\mu\nu}= \hbox{diag}(+1, -1, -1, -1)$ is the metric tensor. In Eq.~(\ref{Tmn}) the vector $h^{\mu}$ represents the heat flux, while $\tau^{\mu\nu}=\pi^{\mu\nu}+\Pi\,\Delta^{\mu\nu}$ is the symmetric dissipative correction to the perfect-fluid form: $\pi^{\mu\nu}$ is the shear stress tensor (the traceless and orthogonal part of $\tau^{\mu\nu}$ related to the shear viscosity) and $\Pi$ is the bulk pressure. In an analogous way, the antisymmetric dissipative corrections are defined by the vector $q^{\mu}$ and the tensor $\phi^{\mu\nu}$. The tensor $S^{\mu\nu}$ in Eq.~(\ref{Slmn}) can be interpreted as the spin density, $S^{\mu\nu}=u_{\lambda}S^{\lambda\mu\nu}$. In Eq.~(\ref{Slmn}) we have also displayed the gradient correction $S^{\lambda\mu\nu}_{(1)}$, which satisfies the constraint $u_{\lambda}S^{\lambda\mu\nu}_{(1)}=0$. We will neglect it below, as it does not contribute to the non-equilibrium entropy current in the order considered in this work~\cite{Hattori:2019lfp}. Finally, we note that the tensors $h^{\mu}$, $\tau^{\mu\nu}$, $q^{\mu}$ and $\phi^{\mu\nu}$ satisfy the following conditions: $h^{\mu}u_{\mu}=0$, $\tau^{\mu\nu}=\tau^{\nu\mu}$, $\tau^{\mu\nu}u_{\nu}=0$, $q^{\mu}u_{\mu}=0$, 
$\phi^{\mu\nu}=-\phi^{\nu\mu}$, and $\phi^{\mu\nu}u_{\nu}=0$. 

Below we will consider the Landau frame by setting $h^{\mu}=0$. We stress that in the presence of an anti-symmetric part of the energy-momentum tensor, the use of the Landau frame is not trivial. In standard hydrodynamics with a symmetric energy-momentum tensor, the Landau frame is defined by the equation $T^{\mu\nu}_{(s)} u_{\nu}=\varepsilon u^{\mu}$. Here $T^{\mu\nu}_{(s)}$ is the symmetric part of the energy-momentum tensor. This implies $h^{\mu}=0$. In the presence of an antisymmetric part, we can also consider the same definition of the Landau frame. However, one may also choose $T^{\mu\nu}u_{\nu}=\varepsilon u^{\mu}$. This would imply $h^{\mu}+q^{\mu}=0$ in the Landau frame. Fortunately, our results do not depend on such choices of the Landau frames, as we argue in a subsequent subsection that for a consistent description of a boost-invariant system we should use $q^{\mu}=0$. We note that the same condition was used in Ref.~\cite{Wang:2021ngp}.

The conservation laws for energy, linear momentum, and angular momentum can be written as
\begin{eqnarray}
(u \cdot \partial) \varepsilon+(\varepsilon+p) \partial \cdot u  \!\!\!&=&\!\!\! -\partial \cdot q
-q^{\nu}(u \cdot \partial) u_{\nu}  \nonumber \\
&& -u_{\nu} \partial_{\mu} \phi^{\mu \nu}
+\tau^{\mu \nu}  \partial_{\mu} u_{\nu},\label{energyeq}\\
(\varepsilon+p) (u \cdot \partial) u^{\alpha} - \Delta^{\alpha\mu} \partial_{\mu} p
\!\!\!&=&\!\!\!-(q \cdot \partial) u^{\alpha}+q^{\alpha}(\partial \cdot u)
\nonumber \\
&&+\Delta^{\alpha}_{~\nu}(u \cdot \partial) q^{\nu}-\Delta^{\alpha}_{~\nu} \partial_{\mu} \phi^{\mu \nu}
\nonumber \\
&& -\Delta^{\alpha}_{~\nu} \partial_{\mu} \tau^{\mu \nu}, 
\label{acceleratioeq} \\
\partial_{\mu}(u^{\mu}S^{\alpha\beta}) 
\!\!\!&=& \!\!\!-2(q^{\alpha}u^{\beta}-q^{\beta}u^{\alpha}+\phi^{\alpha\beta}). 
\label{spineq}
\end{eqnarray}

\subsection{Thermodynamic relations and spin EoS}

Following earlier works~\cite{Hattori:2019lfp,Daher:2022xon,Fukushima:2020ucl}, we assume that the presence of spin degrees of freedom leads to generalized thermodynamic identities:
\begin{align}
    &\varepsilon+p=Ts+\omega_{\alpha\beta}S^{\alpha\beta},\label{eps}\\
    &d\varepsilon=Tds+\omega_{\alpha\beta}dS^{\alpha\beta},\label{deps}\\
    &dp=sdT+S^{\alpha\beta}d\omega_{\alpha\beta}, \label{dP}
\end{align}
where $T$ is the temperature, $s$ is the entropy density, and the anti-symmetric tensor $\omega_{\mu\nu}$ can be interpreted as the spin chemical potential conjugated to the spin density $S^{\mu\nu}$. 

We consider the spin chemical potential to be of the first order in the gradient expansion, i.e., $\omega_{\mu\nu}\thicksim\mathcal{O}(\partial)$. This is implied by the fact that in the presence of the anti-symmetric part of the energy-momentum tensor the quantity $\omega_{\mu\nu}$ can be expressed at global equilibrium in terms of the thermal vorticity tensor~\cite{Hattori:2019lfp}
\begin{equation} 
\omega_{\mu\nu} \rightarrow
\frac{T}{2} \omega^{\rm th}_{\mu\nu} = -\frac{T}{4}
\left(\partial_\mu \beta_\nu - \partial_\nu \beta_\mu \right),
\end{equation}
where $\beta^\mu = \beta u^\mu$ and $\beta=1/T$ is the inverse temperature.\footnote{Note that in the natural units the thermal vorticity is dimensionless, while the spin chemical potential has the mass dimension one.} Consequently, the last terms on the right-hand sides of Eqs.~(\ref{eps})--(\ref{dP}) are not negligible only if the spin density tensor $S^{\alpha\beta}$ is of the zeroth  order in gradients $\mathcal{O}(1)$. In the previous works~\cite{Hattori:2019lfp,Wang:2021ngp} one assumes the form $S^{\mu\nu} \sim T^2 \omega^{\mu\nu}$ and argues that sufficiently large values of $T$ may compensate smallness of $\omega^{\mu\nu}$. We find this argument as not completely convincing since in the hydrodynamic gradient expansion we have $S^{\mu\nu}\sim \mathcal{O}(1)$, $\omega^{\mu\nu}\sim \mathcal{O}(\partial)$, and $T\sim \mathcal{O}(1)$. 

Due to the difficulties outlined above, in this work we propose a different scaling of the spin density tensor, namely, we assume the form
\begin{align}
    S^{\mu\nu}(T,\omega^{\mu\nu})=S_{0}(T)\frac{\omega^{\mu\nu}}{\sqrt{\omega^{\mu\nu}\omega_{\mu\nu}}}\equiv S_{0}(T)\frac{\omega^{\mu\nu}}{\sqrt{\omega:\omega}}.
    \label{spintensor}
\end{align}
This form implies that $S^{\mu\nu}\sim \mathcal{O}(1)$ with $\omega^{\mu\nu}\sim \mathcal{O}(\partial)$. We note that the quantity $\omega:\omega$ is not necessarily positive~\footnote{We expect that solutions of the presented model will split into two categories, one with $\omega:\omega>0$ and the other with $\omega:\omega <0$.}. In the case where it is negative one should include the minus sign within the square root.

Equation (\ref{dP}) implies that pressure can be treated as a function of $T$ and $\omega^{\mu\nu}$. Moreover, Eq.~(\ref{eps}) implies that pressure includes first order corrections in $\omega^{\mu\nu}$. Consequently, we propose a general form of the pressure 
\begin{align}
    p(T,\omega^{\mu\nu})=p_0(T)+p_1(T)\sqrt{\omega:\omega}.
    \label{Peos}
\end{align}
Here $p_0(T)$ is the thermodynamic pressure in the absence of spin chemical potential. Using the thermodynamic relation (\ref{dP}) one obtains
\begin{align}
    S^{\alpha\beta}=\frac{\partial p}{\partial \omega_{\alpha\beta}}\bigg\vert_{T}=p_1(T)\frac{\omega^{\alpha\beta}}{\sqrt{\omega:\omega}}.
    \label{equ14new}
\end{align}
Combing Eqs.~(\ref{spintensor}) and (\ref{equ14new}) we find
\begin{align}
     p_1(T)=S_{0}(T).
     \label{P1}
\end{align}
This relation allows us to write the following expressions for pressure, entropy density, and energy density
\begin{align}
    & p(T,\omega^{\mu\nu})=p_{0}(T)+ S_0(T)\sqrt{\omega:\omega},\label{equ75}\\
     & s(T,\omega^{\mu\nu})=s_{0}(T)+S_0^\prime(T)\sqrt{\omega:\omega},\label{equ76}\\
    & \varepsilon(T,\omega^{\mu\nu})=\varepsilon_{0}(T)+T\,S_0^\prime(T) \sqrt{\omega:\omega},\label{equ77}
\end{align}
where $S_0^\prime(T) = d S_0(T)/d T$. Similarly to $p_0(T)$, the quantities $\varepsilon_{0}(T)$ and $s_{0}(T)$ refer to the case of unpolarized system. They satisfy the thermodynamic relation $\varepsilon_{0}+p_{0}=Ts_{0}$, which is consistent with Eq.~(\ref{eps}).

\subsection{Dissipative currents and kinetic coefficients}

In the last section, we have defined various thermodynamic quantities appearing in the hydrodynamic equations. For a complete description, we also have to specify the dissipative currents appearing in Eqs.~\eqref{energyeq}--\eqref{spineq}. Using the condition of the positive entropy production and restricting our consideration to the linear terms in gradients, one obtains~\cite{Daher:2022xon,Fukushima:2020ucl,Hattori:2019lfp}
\begin{align}
     h^{\mu}  &=-\kappa\left(Du^{\mu}-\beta\nabla^{\mu}T\right), \label{equ3ver2}\\
     q^{\mu}  &=\lambda\left(Du^{\mu}+\beta\nabla^{\mu}T-4\omega^{\mu\nu}u_{\nu}\right),\label{equ4ver2}\\
     \pi^{\mu\nu}&=2\eta_s\sigma^{\mu\nu}, \\
     \Pi&=\zeta\theta, \label{equ5ver2}\\
    \phi^{\mu\nu} 
    &=\gamma(\Omega^{\mu\nu}+2\beta\Delta^{\mu\alpha}\Delta^{\nu\beta}\omega_{\alpha\beta})\nonumber\\ ~~~~~&=\widetilde{\gamma}\left(\nabla^{\mu}u^{\nu}-\nabla^{\nu}u^{\mu}+4\Delta^{\mu\alpha}\Delta^{\nu\beta}\omega_{\alpha\beta}\right).\label{equ6ver2}
\end{align}
Here $\theta\equiv\partial_{\alpha}u^{\alpha}$ is the expansion scalar, $D\equiv u^{\mu}\partial_{\mu}$ is the convective derivative, $\nabla^{\mu}=\Delta^{\mu\nu}\partial_{\nu}$ is the transverse gradient,  $\widetilde{\gamma}=\beta\gamma/2$, $\sigma^{\mu\nu}=\nabla^{(\mu}u^{\nu)}-\frac{1}{3}\theta\Delta^{\mu\nu}$, and $\Omega^{\mu\nu}=\beta\nabla^{[\mu}u^{\nu]}$. Note that $h^{\mu}$, $q^{\mu}$, $\tau^{\mu\nu}$ and $\phi^{\mu\nu}$ are all $\mathcal{O}(\partial)$ in the hydrodynamic gradient expansion. All transport coefficients are positive, i.e., $\kappa\geq 0$, $\lambda\geq 0$, $\eta_s\geq 0$, $\zeta\geq 0$, and $\gamma\geq 0$, to ensure positive entropy production in a dissipative system.  With the specified equation of state and dissipative currents, the system of hydrodynamic equations becomes closed and we can search for its solutions. 

\section{Boost-invariant description of spin hydrodynamics}
\label{sec:implementBI}
\subsection{Implementation of the boost invariance}

For the boost-invariant systems which are uniform in the transverse plane, the hydrodynamic flow has the form $u^{\mu}\!=\!(\cosh\eta,0,0,\sinh\eta)$, where $\eta =(1/2)\ln[(t+z)/(t-z)]$ is the spacetime rapidity~\cite{Bjorken:1982qr}. Moreover, all thermodynamic quantities  depend only on the proper time $\tau = \sqrt{t^2-z^2}$. In this case, the four-acceleration of the fluid, $a^\mu = Du^{\mu}$, as well as the transverse gradient of temperature, $\nabla^{\mu}T$, vanish. Consequently, the tensors $q^{\mu}$ and $\phi^{\mu\nu}$ defined by Eqs.~\eqref{equ4ver2} and \eqref{equ6ver2} can be directly expressed by the spin chemical potential~\cite{Wang:2021ngp}
\begin{align}
q^{\mu} &=-4 \lambda\, \omega^{\mu \nu} u_{\nu} \label{equ24new},\\
\phi^{\mu \nu} &=2 \gamma\beta\left(\omega^{\mu \nu}+2 u^{[\mu} \omega^{\nu] \beta} u_{\beta}\right).\label{equ25new}
\end{align}
While dealing with boost-invariant system, it is also convenient to introduce the following basis vectors:
\begin{align}
 u^{\mu}&\equiv(\cosh\eta,0,0,\sinh\eta),\\
 X^{\mu}&\equiv (0,1,0,0),\\
 Y^{\mu}&\equiv (0,0,1,0),\\
 Z^{\mu}&\equiv (\sinh\eta,0,0,\cosh\eta).
\end{align}
The fluid flow four-vector $u^{\mu}$ is time-like, while the four-vectors $X^{\mu}$, $Y^{\mu}$, and $Z^{\mu}$ are space-like and orthogonal to $u^{\mu}$. They satisfy the following properties:
\begin{align}
    & 
u^{\mu}X_{\mu}=0, \quad u^{\mu}Y_{\mu}=0, \quad u^{\mu}Z_{\mu}=0,\\
    & X^{\mu}X_{\mu}=-1, \quad Y^{\mu}Y_{\mu}=0, \quad Z^{\mu}Z_{\mu}=0,\\
    & X^{\mu}Y_{\mu}=-1, \quad X^{\mu}Z_{\mu}=0,\quad Y^{\mu}Z_{\mu}=-1. 
\end{align}
The spin chemical potential $\omega^{\mu\nu}$ is anti-symmetric and can be generally decomposed as~\cite{Florkowski:2017ruc}
\begin{align}
    \omega^{\mu\nu}=\kappa^{\mu}u^{\nu}-\kappa^{\nu}u^{\mu}+\epsilon^{\mu\nu\alpha\beta}u_{\alpha}\omega_{\beta}.
    \label{equ15}
\end{align}
Here, the four vectors $\kappa^{\mu}$ and $\omega^{\mu}$ are also space-like and orthogonal to $u^{\mu}$, i.e., $\kappa^{\mu}u_{\mu}=0$ and $\omega^{\mu}u_{\mu}=0$.  Using the decomposition~\eqref{equ15} in Eqs.~\eqref{equ24new}--\eqref{equ25new}, we find %
\begin{align}
    & q^{\mu}=-4\lambda\kappa^{\mu},\label{equ35new}\\
    & \phi^{\mu\nu}=2\beta\gamma\epsilon^{\mu\nu \alpha \beta}u_{\alpha}\omega_\beta.
    \label{equ36new}
\end{align}
Furthermore, using the space-like basis vectors $X^{\mu}$, $Y^{\mu}$, and $Z^{\mu}$, one can introduce the following representation of the vectors $\kappa^{\mu}$ and $\omega^{\mu}$~\cite{Florkowski:2019qdp}
\begin{align}
    \kappa^{\mu} & \equiv C_{\kappa X}X^{\mu}+C_{\kappa Y}Y^{\mu}+C_{\kappa Z}Z^{\mu}\nonumber\\
    & =\left(C_{\kappa Z}\sinh\eta,C_{\kappa X},C_{\kappa Y},C_{\kappa Z}\cosh\eta\right),\label{equ37new}\\
    \omega^{\mu} & \equiv C_{\omega X}X^{\mu}+C_{\omega Y}Y^{\mu}+C_{\omega Z}Z^{\mu}\nonumber\\
    & =\left(C_{\omega Z}\sinh\eta,C_{\omega X},C_{\omega Y},C_{\omega Z}\cosh\eta\right).\label{equ38new}
\end{align}

\subsection{Boost-invariant fire-cylinder: spin and orbital angular momentum}

In order to get a physical insight into the coefficients $C_{\kappa i}$, and $C_{\omega i}$, we calculate the spin and orbital angular momentum of the fire-cylinder (FC) occupying the space-time region defined by the conditions: $\tau$=const., $-\eta_{\rm FC}/2 \leq \eta \leq \eta_{\rm FC}/2$, and $\sqrt{x^{2}+y^{2}}\leq R$ (see also Fig.~1 in Ref.~\cite{Florkowski:2019qdp}). A space-time volume element of the fire-cylinder can be defined as
\begin{equation}
    d\Sigma_{\lambda}=u_{\lambda}dxdy\tau d\eta.
\end{equation}
Using the leading term of the spin tensor, one can calculate the spin angular momentum contained in the fire-cylinder
\begin{align}
    \mathcal{S}^{\mu\nu}_{\rm FC}&=\int_{}^{} d\Sigma_{\lambda}~S^{\lambda\mu\nu}_{\rm ph}
    =\int_{}^{} dxdy\tau d\eta~ S^{\mu\nu}_{\rm ph}\nonumber\\
    &=\int_{}^{} dxdy\tau d\eta\frac{S_{0}}{\sqrt{\omega:\omega}}\nonumber\\
    &\quad \quad \quad \times(\kappa^{\mu}u^{\nu}-\kappa^{\nu}u^{\mu}+\epsilon^{\mu\nu\alpha\beta}u_{\alpha}\omega_{\beta}).\label{totalspin}
\end{align}
With the help of Eq.~(\ref{totalspin}), the $\mathcal{S}^{0i}_{\rm FC}$ components can be obtained, which read:
\begin{align}
    &\mathcal{S}^{01}_{\rm FC}=-2\pi R^{2}\tau \frac{S_{0}}{\sqrt{ \omega:\omega}}C_{\kappa X}\sinh(\frac{\eta_{\rm FC}}{2}),\\
    &\mathcal{S}^{02}_{\rm FC}=-2\pi R^{2}\tau\frac{S_{0}}{\sqrt{ \omega:\omega}}C_{\kappa Y}\sinh(\frac{\eta_{\rm FC}}{2}),\\
    &\mathcal{S}^{03}_{\rm FC}=-\pi R^{2}\tau\frac{S_{0}}{\sqrt{ \omega:\omega}}C_{\kappa Z}\eta_{\rm FC}.
\end{align}
On the other hand, the orbital part of the total angular momentum can be obtained from the formula
\begin{align}
\label{totalobital}
    \mathcal{L}^{\mu\nu}_{\rm FC}=\int_{}^{} d\Sigma_{\lambda}L^{\lambda\mu\nu}_{\rm ph}=\int_{}^{} d\Sigma_{\lambda}\left(x^{\mu}T^{\lambda\nu}_{\rm ph}-x^{\nu}T^{\lambda\mu}_{\rm ph}\right),
\end{align}
which leads to the expression
\begin{align}
    \mathcal{L}^{\mu\nu}_{\rm FC}= \int~dx dy \tau d\eta\left(\varepsilon x^{\mu}u^{\nu}-\varepsilon x^{\nu}u^{\mu}-x^{\mu}q^{\nu}+x^{\nu}q^{\mu}\right).
\end{align}
In the above equation, we have used the orthogonality condition between the fluid four-velocity and various dissipative currents. Using the representation of $q^{\mu}$ in terms of $\kappa^{\mu}$ as given in Eq.~\eqref{equ35new}, we obtain the $\mathcal{L}^{0i}_{\rm FC}$ components, 
\begin{align}
    &\mathcal{L}^{01}_{\rm FC}=8\lambda C_{\kappa X}\pi R^{2}\tau^{2}\sinh(\frac{\eta_{\rm FC}}{2}),\\
    &\mathcal{L}^{02}_{\rm FC}=8\lambda C_{\kappa Y}\pi R^{2}\tau^{2}\sinh(\frac{\eta_{\rm FC}}{2}),\\
    &\mathcal{L}^{03}_{\rm FC}=4\lambda C_{\kappa Z}\pi R^{2}\tau^{2}\eta_{\rm FC}.
\end{align}
Interestingly, the coefficients $C_{\omega i}$ do not appear in the equations above.

The ${\cal J}^{0i}_{\rm FC}\equiv\mathcal{L}^{0i}_{\rm FC}+\mathcal{S}^{0i}_{\rm FC}$  components of the fire-cylinder describe its center-of-mass motion and should vanish in the center-of mass system that we use here. The conditions ${\cal J}^{0i}_{\rm FC}=0$ (for $i=1,2,3$) can be explicitly rewritten as:
\begin{align}
    &\pi R^2\tau ~C_{\kappa X}\left(- \frac{S_{0}}{\sqrt{ \omega:\omega}}+4\lambda\tau\right)\sinh(\frac{\eta_{\rm FC}}{2})=0,\label{con1}\\
    &\pi R^2\tau ~C_{\kappa Y}\left(- \frac{S_{0}}{\sqrt{ \omega:\omega}}+4\lambda\tau\right)\sinh(\frac{\eta_{\rm FC}}{2})=0,\label{con2}\\
& \pi R^2\tau ~C_{\kappa Z} \left(- \frac{S_{0}}{\sqrt{ \omega:\omega}}+4\lambda\tau\right)\eta_{\rm FC} =0.\label{con3}
\end{align}
Since $\lambda$ and $S_0$ are independent quantities, Eqs.~\eqref{con1}--\eqref{con3} can be fulfilled only if the coefficients $C_{\kappa i}$ vanish,
\begin{equation}
    C_{\kappa X}=C_{\kappa Y}=C_{\kappa Z}=0.\label{solution}
\end{equation}
The solution (\ref{solution}) also implies that $\kappa^{\mu}=0$, and consequently $q^{\mu}=0$. Moreover, in this case $\omega^{\mu\nu}$ is determined entirely by $\omega^{\mu}$ and
\begin{align}
\omega^{\mu\nu}\omega_{\mu\nu}&=-2\omega^{\mu}\omega_{\mu}\nonumber\\
&=2(C_{\omega X}^2+C_{\omega Y}^2+C_{\omega Z}^2)\nonumber\\
&\equiv 2C^{2}>0.
\label{equ54new}
\end{align}
In the absence of $q^{\mu}$, the orbital part of the fire-cylinder becomes
\begin{align}
    \mathcal{L}^{\mu\nu}_{\rm FC}= \int \varepsilon \left( x^{\mu}u^{\nu}- x^{\nu}u^{\mu}\right)~dx dy \tau d\eta.
\end{align}
Using the explicit form of the fluid four-velocity for the Bjorken flow in the above equation and performing the space-time integration, it can be argued that the orbital angular momentum of the fire-cylinder vanishes,  $\mathcal{L}^{\mu\nu}_{\rm FC}~=~0$~\cite{Florkowski:2019qdp}. 

\subsection{Hydrodynamic equations for a boost-invariant system}

Using the Landau frame as specified above, along with the condition $q^{\mu}=0$ and the explicit expression for $\phi^{\mu\nu}$ as given by Eq.~\eqref{equ36new}, one obtains the energy conservation equation for a boost-invariant system,   
\begin{align}
    \frac{d\varepsilon}{d\tau}+\frac{\varepsilon+p}{\tau}-\frac{1}{\tau}\bigg(\frac{2}{3}\frac{\eta_s}{\tau}+\frac{\zeta}{\tau}\bigg)=0.
\end{align}
Although $\phi^{\mu\nu}$ explicitly appears in Eq.~\eqref{energyeq},  it can be shown for the Bjorken flow that $u_{\nu}\partial_{\mu}\phi^{\mu\nu}=0$. Therefore, the antisymmetric parts of the energy-momentum tensor do not explicitly appear in the evolution of energy density (note, however, that they appear implicitly as they affect the constitutive equations (\ref{equ75})--(\ref{equ77})). 

Note that $\eta_s/s_0$ and $\zeta/s_0$ are dimensionless quantities and in the hydrodynamic description they can play an important role. In terms of these dimensionless variables  the above equation can be rewritten as
\begin{align}
    \frac{d\varepsilon}{d\tau}+\frac{\varepsilon+p}{\tau}-\frac{s_0}{\tau^2}\bigg(\frac{2}{3}\frac{\eta_s}{s_0}+\frac{\zeta}{s_0}\bigg)=0.
    \label{equ26}
\end{align}
Let us now consider the acceleration equation (\ref{acceleratioeq}). For the Bjorken flow we can use the properties:
\begin{align}
    & (u\cdot\partial) u^{\mu}=0,\\
    & \Delta^{\alpha\mu}\partial_{\mu}p=0,\\
    & \Delta^{\alpha}_{~\nu}\partial_{\mu}\phi^{\mu\nu}=0,\\
    & \Delta^{\alpha}_{~\nu}\partial_{\mu}\tau^{\mu\nu}=0.
\end{align}
Therefore, for the case $q^{\mu}=0$, the acceleration equation (\ref{acceleratioeq}) is trivially fulfilled. 

The remaining equation that has to be considered in our scheme is the spin evolution equation (\ref{spineq}). Note that the antisymmetric part of the energy-momentum tensor cannot be neglected in this case. However, in the case $q^{\mu}=0$, only $\phi^{\mu\nu}$ affects the spin evolution equation, namely
\begin{align}
    \frac{\partial S^{\mu\nu}}{\partial\tau}+\frac{S^{\mu\nu}}{\tau}=-2\phi^{\mu\nu}.
    \label{Smn}
\end{align}
Note that both the left-hand side as well as the right-hand side of the above equation depend on $\omega^{\mu\nu}$. Since $\kappa^{\mu}=0$, the spin chemical potential $\omega^{\mu\nu}$ is completely determined by the four-vector $\omega^{\mu}$ which has three independent components (due to the orthogonality condition $\omega^{\mu}u_{\mu}=0$). Therefore, Eq.~(\ref{Smn}) can be rewritten as three differential equations governing the evolution of three independent components of the tensor $\omega^{\mu\nu}$. These equations can be obtained by taking projections of the spin evolution equation~(\ref{Smn}) with $X_{\mu}Y_{\nu}$, $X_{\mu}Z_{\nu}$, and $Y_{\mu}Z_{\nu}$. This gives us the following set of equations for the coefficients $C_{\omega X}$, $C_{\omega Y}$, and $C_{\omega Z}$ 

\begin{align}
    & \frac{d}{d\tau}\left(\frac{S_0}{\sqrt{2}C}C_{\omega X}\right)+\left(\frac{S_0}{\sqrt{2}C}C_{\omega X}\right)\frac{1}{\tau}=-4\beta\gamma C_{\omega X}, \label{equ68}\\
    & \frac{d}{d\tau}\left(\frac{S_0}{\sqrt{2}C}C_{\omega Y}\right)+\left(\frac{S_0}{\sqrt{2}C}C_{\omega Y}\right)\frac{1}{\tau}= -4\beta\gamma  C_{\omega Y}, \label{equ69}\\
    & \frac{d}{d\tau}\left(\frac{S_0}{\sqrt{2}C}C_{\omega Z}\right)+\left(\frac{S_0}{\sqrt{2}C}C_{\omega Z}\right)\frac{1}{\tau}=-4\beta\gamma C_{\omega Z},\label{equ70}
\end{align}
where $C = \sqrt{C_{\omega X}^2+C_{\omega Y}^2+C_{\omega Z}^2}$, see Eq.~\eqref{equ54new}. After several  algebraic manipulations Eqs.~\eqref{equ68}--\eqref{equ70} yield
\begin{eqnarray}
    C &=&-\frac{1}{4\sqrt{2}\gamma\beta}\left(\frac{dS_{0}}{d\tau}+\frac{S_{0}}{\tau}\right) \nonumber \\
    &=&-\frac{1}{4\sqrt{2}\gamma\beta}\left(S_0^\prime(T) \frac{dT}{d\tau}+\frac{S_{0}}{\tau}\right).
    \label{spinevolution}
\end{eqnarray}
To obtain the above equation we have assumed that $C~\neq~0$.

In order to get some insight into Eqs.~(\ref{equ68})--(\ref{equ70}), we express different components of the spin chemical potential in terms of $C$ and two spherical angles $(\Theta,\Phi)$:
\begin{align}
    & C_{\omega X}= C\sin\Theta \cos\Phi,\\
    & C_{\omega Y}= C\sin\Theta \sin\Phi,\\
    & C_{\omega Z}= C\cos\Theta.
\end{align}
In this case, Eq.~\eqref{equ70} can be written as  
\begin{align}
   & \frac{d}{d\tau}\left(\frac{S_0}{\sqrt{2}}\cos\Theta\right)+\left(\frac{S_0}{\sqrt{2}}\cos\Theta\right)\frac{1}{\tau}=-4\beta\gamma C \cos\Theta
\end{align}
which implies
\begin{align}
& - \frac{S_0}{\sqrt{2}}\sin\Theta~\frac{d\Theta}{d\tau}+\frac{1}{\sqrt{2}}\cos\Theta\left(\frac{dS_{0}}{d\tau}+\frac{S_{0}}{\tau}\right)\nonumber\\
  & ~~~~~~~~~~~~~~~~~~~~~~~~~~~~~~~~~~=-4\beta\gamma C \cos\Theta
\end{align}
and, consequently, $d\Theta/d\tau=0$. To obtain the last condition we used Eq.~\eqref{spinevolution}. Similarly, using Eq.~\eqref{equ68} or \eqref{equ69}, it can be easily shown that $d\Phi/d\tau=0$. Therefore, the vector built out of three components $C_{\omega X}$, $C_{\omega Y}$, and $C_{\omega Z}$ does not change its direction during the time evolution (its orientation is fixed by the initial condition) and the only non-trivial dependence is that of the magnitude $C$.   

In this way, our analysis has been reduced to a study of  the hydrodynamic equation~(\ref{equ26}), where
\begin{align}
    & p(T,C)=p_{0}(T)+ \sqrt{2} S_0 C,\label{equ75a}\\
    & \varepsilon(T,C)=\varepsilon_{0}(T)+\sqrt{2}TS_0^\prime(T) C,\label{equ77a}
\end{align}
and the spin evolution equation (\ref{spinevolution}). We note that Eq.~\eqref{equ26} is a first-order differential equation for the energy density. By the virtue of Eq.~\eqref{equ77}, the energy density depends on both the temperature and the spin chemical potential, i.e., it is a function of $T$ and $C$. Therefore, one can combine  Eqs.~\eqref{equ26} and \eqref{spinevolution} into a single second-order differential equation for the proper time evolution of $T$. Once this equation is solved with appropriate initial conditions, we can obtain a proper time evolution of temperature. Subsequently, we obtain the proper time dependence of $C$ using Eq.~\eqref{spinevolution}. 

As in standard hydrodynamic models, the procedure outlined above works in practice if the equation of state is known. In our case, with spin degrees of freedom included, this means that we have to know the temperature dependence of the function $S_0$.      


\section{Spin equation of state and transport coefficient $\gamma$}
\label{sec:EOS}
%
In Eq.~\eqref{spintensor}, we have introduced a relation between the spin density tensor $S^{\mu\nu}$ and the spin chemical potential $\omega^{\mu\nu}$. To completely specify the spin equation of state, we need to know the form of the temperature-dependent function $S_0(T)$. Unfortunately, at the moment we lack any microscopic models for this function. Consequently, in our numerical calculations, we have used arguments that refer to dimensional analysis and overall simplicity. Since $S_0(T)$ has mass-dimension three, we have explored the form
\begin{align}
   S_{0}(T)=\frac{\alpha}{\sqrt{2}}T^{a}M^{b}K_{n}\left(\frac{M}{T}\right),\label{s0}
\end{align}
where $a$ and $b$ are numerical constants satisfying the condition $a+b=3$. Here $M$ is the particle mass, and $K_{n}$ denotes the modified Bessel function of the second kind. The parameter $\alpha$ is a numerical constant. The appearance of the modified Bessel function in Eq.~(\ref{s0}) is connected with the fact that we are going to include, as special cases, the scaling of the function $S_0(T)$ with particle density $n_0(T)$ or entropy density $s_0(T)$. These two cases can be obtained by the appropriate choice of $\alpha$, $a$, and $n$ in Eq.~\eqref{s0}. 

Using Eq.~\eqref{s0}, the thermodynamic variables defined by Eqs.~\eqref{equ75}--\eqref{equ77} can be expressed in the following manner:
\begin{align}
    &p(T,C)=p_{0}(T)+\alpha T^a M^b K_n\left(\frac{M}{T}\right)~C,\label{pressureeq}\\
    &s(T,C)=s_{0}(T)+\alpha aT^{a-1}M^b K_n\left(\frac{M}{T}\right)C\nonumber\\
    &~~~~~~~~~~~~~~~~~+\alpha T^a M^b K_n^{\prime}\left(\frac{M}{T}\right)C,\\
    &\varepsilon(T,C)=\varepsilon_{0}(T)+a\alpha T^a M^b K_n^{}\left(\frac{M}{T}\right)~C\nonumber\\
    &~~~~~~~~~~~~~~~~~+\alpha T^{a+1} M^b K_n^{\prime}\left(\frac{M}{T}\right)~C,
    \label{energydenistyeq}
\end{align}
where the prime in  $K_n^{\prime}\left(M/T\right)$ denotes a derivative with respect to temperature. For a massive Boltzmann gas, the explicit expressions for thermodynamic quantities without the spin part are %
\begin{align}
    & p_{0}=\frac{g_s}{2\pi^{2}}T^{2}M^{2}K_{2}\left(\frac{M}{T}\right),\label{pressure0}\\
    & \varepsilon_{0}=\frac{g_s}{2\pi^{2}}T^{2}M^{2}\bigg[3K_{2}\left(\frac{M}{T}\right)+\frac{M}{T}K_{1}\left(\frac{M}{T}\right)\bigg],\\
    & n_0=p_0/T,\\
    & s_0=\frac{(\varepsilon_0+p_0)}{T}=\frac{g_s}{2\pi^2}M^3K_3\left(\frac{M}{T}\right),
\end{align}
where $g_s$ is the spin and particle-antiparticle degeneracy factor.

To obtain the proper time evolution of temperature and spin chemical potential, we need to specify the transport coefficient $\gamma$, since it is present in Eq.~\eqref{spinevolution}.
To select a specific form of $\gamma$ we use again dimensional analysis. In analogy to Eq.~\eqref{s0}, we use
\begin{align}
    \gamma=\tilde{\alpha} \, T^{c+1} M^{d} K_{m}\left(\frac{M}{T}\right)\label{gamma}.
\end{align}
where $\tilde{\alpha}$, $c$, $d$, and $m$ are numerical constants, with \mbox{$c+d=3$}. To demonstrate the effect of different forms of $S_0(T)$ and $\gamma$ on the proper time evolution of temperature and spin chemical potential, we consider two different options: 

\medskip
{\bf Case I}:
In this case, we assume that the function $S_0(T)$ is given by the particle density $n_0(T)$, while $\gamma(T)$ is proportional to the pressure $p_0(T)$,
\begin{align}
    & S_0(T)\equiv n_0(T), ~~ \gamma(T) \equiv \mathcal{A}\,p_0(T).
\end{align}
These expressions for $S_0(T)$ and $\gamma(T)$ can be obtained from Eqs.~\eqref{s0} and \eqref{gamma} by choosing: $a=1$, $n=2$, $\alpha = g_s \sqrt{2}/(2\pi^2)$, $c=1$, $m=2$, and $\mathcal{A}= 2\pi^{2} \tilde{\alpha}/g_s$. In order to be able to treat the spin effects as a small correction to the standard (spinless) dynamics, we assume a very small value of the parameter $\tilde{\alpha}$. In practice we choose $\tilde{\alpha}=0.001$.  

{\bf Case II}: In this case, we assume that the functions $S_0(T)$ and $\gamma(T)$ are closely related to the entropy density $ s_0(T)$, namely
\begin{align}
    & S_0(T)\equiv s_0(T), ~~ \gamma(T) \equiv T s_0(T).
\end{align}
These expressions for $S_0(T)$ and $\gamma(T)$ can be obtained from Eqs.~\eqref{s0} and \eqref{gamma} by choosing:  $a=0$, $n=3$, $\alpha=g_s\sqrt{2}/(2\pi^2)$,  $c=0$,  $m=3$, and $\tilde{\alpha}=g_s/(2\pi^2)$. 

\section{Numerical results}
\label{sec:numerics}

If the functions $S_0(T)$ and $\gamma(T)$ are defined by Eqs.~\eqref{s0} and \eqref{gamma}, respectively, the hydrodynamic evolution equations for the proper-time dependence of temperature and the magnitude of spin chemical potential, Eqs.~\eqref{equ26} and~\eqref{spinevolution}, can be written in a compact form as
\begin{align}
 &A(\tau) \frac{d^2T}{d\tau^2}+B(\tau)\bigg(\frac{dT}{d\tau}\bigg)^2+D(\tau)\frac{dT}{d\tau}+E(\tau)=0,
    \label{equ88new} \\
    & C(\tau)=C_{11}(\tau)\frac{dT}{d\tau}+C_{12}(\tau).\label{equ77new}
\end{align}
The explicit expressions for various coefficient functions appearing above are given in Appendix \ref{appendix1}.   
Equation~\eqref{equ88new} is a second-order ordinary differential equation governing the proper-time evolution of temperature. A unique solution of Eq.~\eqref{equ88new} can be found if the initial conditions are specified at $\tau=\tau_0$ for the functions $T(\tau)$ and $dT(\tau)/d\tau$. We note that the initial value of the temperature gradient can be obtained from Eq.~\eqref{equ77new} if the initial values for $T(\tau)$ and $C(\tau)$ are known. Hence, as expected, the full dynamics of our system is determined by the initial values of the temperature and magnitude of the spin chemical potential. We also note that Eq.~\eqref{equ88new} has the form of the Riccati equation that may have analytic solutions for some specific choices of the coefficients, however, in general, one has to solve it numerically. 

In this work, we present our numerical solutions obtained for the two cases defined in the previous section. We assume the initial temperature $T_{0}=T(\tau_0)=200$~MeV at  $\tau_0=0.5$ fm and the initial magnitude of the spin chemical potential $C_{0}=C(\tau_0)=50$~MeV. The internal degeneracy factor equals $g_s=4.0$ (particles and antiparticles with spin 1/2), and $M=500$ MeV represents an effective particle mass. Moreover, we use the result $\eta_s/s_0=1/(4\pi)$, and ignore the effect of the bulk viscosity.

\begin{figure}[h!   ]
	\includegraphics[scale=0.5]{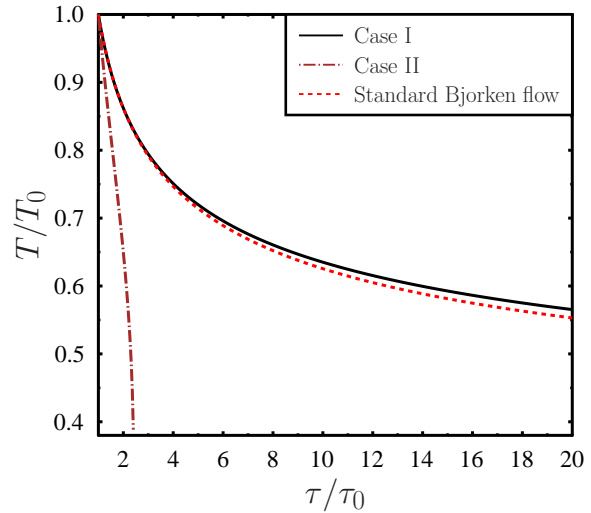}
	\caption{Proper time evolution of temperature. Black (solid) line represents the temperature evolution for case I. Brown (dashed-dotted) line represents the temperature evolution for case II. Red (dashed) line represents the variation of temperature with the proper time for the standard Bjorken flow without spin. We consider $T_0=200$ MeV and $\tau_0=0.5$ fm.   }
	\label{fig:1}
\end{figure}

In Fig.~\ref{fig:1} we present the proper-time evolution of temperature. The solid (black) line and the dashed-dotted (brown line) represent the solutions of Eq.~\eqref{equ88new} for cases I and II, respectively, while the dashed (red) line represents the standard Bjorken flow solution (spinless fluid). Note that for the standard Bjorken flow the temperature evolution equation is a first-order differential equation. We emphasize that one cannot set $C=0$ in Eq.~\eqref{equ88new} to obtain the standard Bjorken flow without spin, since to obtain Eq.~\eqref{equ88new} we used Eq.~\eqref{spinevolution} which assumes $C\neq0$. 

As the temperature profile for the evolution with spin is very close to the standard Bjorken solution in case I, we observe a rapid drop of temperature for case II. For larger evolution times, temperature becomes negative for case II, which suggests that this solution cannot be accepted as physically meaningful. 

\begin{figure}[]
	\includegraphics[scale=0.5]{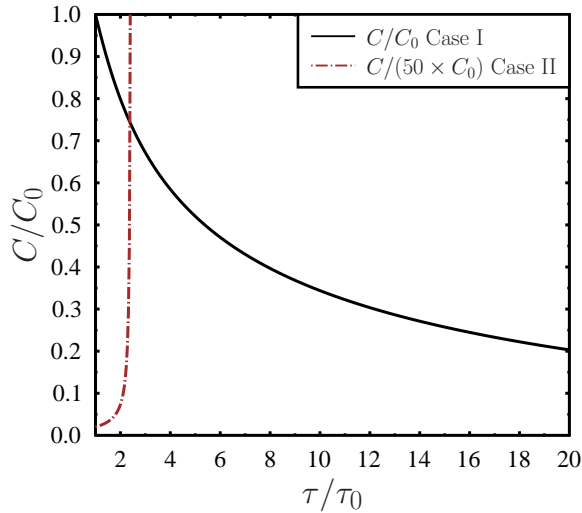}
	\caption{Proper time evolution of the magnitude of the spin chemical potential $C$.  We consider $C_0=50$ MeV and $\tau_0=0.5$ fm. For case I the $C$ decreases with proper time. But for the case II the spin chemical potential grows rapidly with proper time.   }
	\label{fig:2}
\end{figure}
In Fig.~\ref{fig:2} we present the proper-time evolution of the magnitude of the spin chemical potential $C(\tau)$. We clearly observe that $C$ decreases with proper time for case I. This behavior could be interpreted as an approach of spin chemical potential to  its global-equilibrium value given by thermal vorticity that vanishes for Bjorken flow. However, the Killing condition  $\partial_{\mu}\beta_{\nu}+\partial_{\nu}\beta_{\mu}=0$ is not satisfied by the boost invariant flow, hence, strictly speaking, the global equilibrium can never be reached in the considered case. 

Again, for case II we observe a singular behavior, in this case the spin chemical potential rapidly grows with time. Clearly, the assumed equation of state and form of the $\gamma$ coefficient for case II lead to unphysical behavior of the system.

\begin{figure}[]
	\includegraphics[scale=0.5]{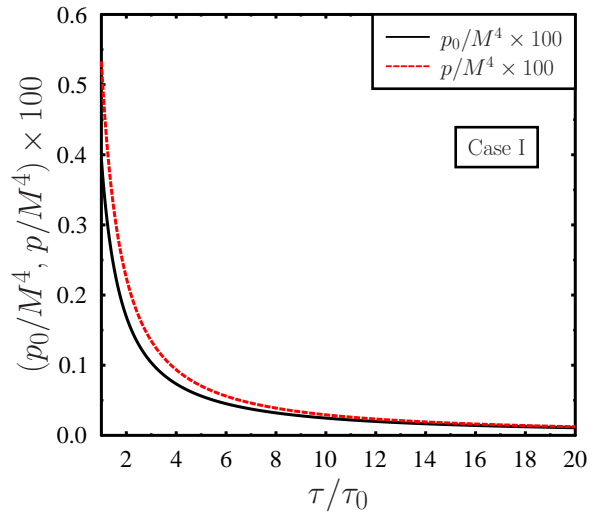}
	\caption{Proper time evolution of $p_0(T)$, and $p(T,C)$ for case I. We observe that the spin contribution to pressure remains a small correction to the total pressure during the whole evolution of the system.  }
	\label{fig:3}
\end{figure}

Finally, in Fig.~\ref{fig:3} we demonstrate the spin contribution to the thermodynamic pressure. The solid (black) line  represents the pressure without the spin contribution, i.e., the function $p_0(T)$. Note that the temperature $T(\tau)$ is the solution of the coupled equations \eqref{equ88new} and \eqref{equ77new}. The dashed (red) line represents the total pressure $p(T,C)$ including the effect of the spin chemical potential. With our choice of parameters, the difference between $p$ and $p_0$ remains small during the whole time evolution of the system. At the end of the evolution, when the value of $C$ is small, the difference between $p_0$ and $p$ is almost negligible.  This indicates that our expansion scheme is consistently realized by the solution in case I.

\section{Conclusion}
\label{sec:conclusions}

In this work we have analyzed the boost-invariant version of the spin hydrodynamic equations formulated by Hattori et al. in Ref.~\cite{Hattori:2019lfp}. A novel feature of our treatment was a form of the spin density tensor that was manifestly of the first order in gradients. We have numerically solved the resulting system of differential equations and found that they exhibit both stable and unstable behavior, depending on the assumed spin equation of state and form of the spin kinetic coefficient $\gamma$. Our finding of unstable behavior may be connected with other results that indicate a similar behavior of spin hydrodynamics if one restricts oneself to only first-order terms in gradients. Clearly, the standard relativistic  Navier-Stokes theory also has  problems with stability and causality --- the fact that triggered development of the second-order Israel-Stewart framework. Our results indicate that a similar extension for spin hydrodynamics is also necessary~\cite{Weickgenannt:2022zxs}.

Some of the problems encountered in our approach may be due to the assumed boost invariance. As we have mentioned above, the Killing equation is never satisfied for such a geometry, hence, the boost-invariant systems cannot reach a state of global thermodynamic equilibrium. Boost invariance imposes also many constraints on the dynamic quantities, which make the spin evolution rather trivial, in particular, no mixing between different components of the spin density is possible. 

Clearly, other forms of the spin equation of state and the spin kinetic coefficient $\gamma$ could be analyzed in the context of stability of the hydrodynamic solutions. However, the two cases analyzed herein correspond most likely to two extreme situations one can encounter.  

The main conclusion of the present work and other recent studies of stability of spin hydrodynamics~\cite{Daher:2022wzf,Sarwar:2022yzs} is the need of constructing of the second order theory that may be free of the problems discussed above. 

\medskip 
{\bf Acknowledgements:} RB thanks the Institute of Theoretical Physics of the Jagiellonian University for financial support organized within the NISER (Bhubaneswar, India) -- Jagiellonian University (Kraków, Poland) Academic Cooperation Agreement. RB also acknowledges the financial support from SPS, NISER planned project RIN 4001. This work was supported in part by the Polish National Science Centre Grant No. 2018/30/E/ST2/00432.

\appendix

\section{}
\label{appendix1}
Using Eqs.~\eqref{s0} and \eqref{gamma} back into Eq.~\eqref{spinevolution} and considering that for the Bjorken flow, $T$ is only a function of the proper time $(\tau)$, we find, 
\begin{equation}
     C=C_{11}\frac{dT}{d\tau}+C_{12},
\end{equation}
\begin{equation}
     C_{11} = -\frac{\alpha}{8\tilde{\alpha}}T^{a-c}M^{c-a}\left(a  \frac{K_n}{T K_m}+\frac{K_n^{\prime}}{K_m}\right),
\end{equation}

\begin{equation}
     C_{12} = -\frac{\alpha}{8\tilde{\alpha}}T^{a-c}M^{c-a}\frac{1}{\tau}\frac{K_n}{K_m}.
\end{equation}
In the above equation, the argument of the Bessel functions is $M/T$. But for a neat representation, we have removed the arguments of the Bessel functions. Moreover one can write, 
\begin{align}
    \frac{dC}{d\tau}=C_{22}\frac{d^2T}{d\tau^2}+C_{21}\left(\frac{dT}{d\tau}\right)^2+C_{23}\left(\frac{dT}{d\tau}\right)+C_{24}.
    \label{equ80new}
\end{align}
Various temperature-dependent coefficients appearing in the above equation are given as,
\begin{align}
    & C_{22} = -\frac{\alpha}{8\tilde{\alpha}}T^{a-c}M^{c-a}\bigg[a  \frac{K_n}{T K_m}+\frac{K_n^{\prime}}{K_m}\bigg], 
\end{align}
\begin{align}
    & C_{21} = -\frac{\alpha}{8\tilde{\alpha}}M^{c-a}\bigg[a(a-c-1)T^{a-c-2}\frac{K_n}{K_m}\nonumber\\
    & ~~~~~~~~+aT^{a-c-1}\frac{K_n^{\prime}}{K_m}-aT^{a-c-1}\frac{K_nK_m^{\prime}}{K_m^2}\nonumber\\
    & +(a-c)T^{a-c-1}\frac{K_n^{\prime}}{K_m}+T^{a-c}\frac{K_n^{\prime\prime}}{K_m}-T^{a-c}\frac{K^{\prime}_m K^{\prime}_n}{K_m^2}\bigg],
\end{align}

\begin{align}
    & C_{23} = -\frac{\alpha}{8\tilde{\alpha}}T^{a-c}M^{c-a}\frac{1}{\tau}\bigg[(a-c) \frac{K_n}{T K_m}+ \frac{K_n^{\prime}}{K_m}\nonumber\\ 
    & ~~~~~~~~~~~~~~~~~~~~~~~~~~~~~~~-\frac{K_n K_m^{\prime}}{K_m^2}\bigg],\\
    & C_{24} = \frac{\alpha}{8\tilde{\alpha}}T^{a-c
    }M^{c-a} \frac{1}{\tau^2}\frac{K_n}{K_m}.
\end{align}
Using the expression of energy density $(\varepsilon)$, and $dC/d\tau$ as given in Eqs.~\eqref{energydenistyeq} and \eqref{equ80new}, respectively, we obtain, 
\begin{align}
    \frac{d\varepsilon}{d\tau}& =\varepsilon_2C_{22}\frac{d^2T}{d\tau^2}+(\varepsilon_2C_{21}+\varepsilon_1 C_{11})\left(\frac{dT}{d\tau}\right)^2\nonumber\\
    & +\bigg(\frac{d\varepsilon_{0}}{dT}+C_{12}\varepsilon_1+C_{23}\varepsilon_2\bigg)\left(\frac{dT}{d\tau}\right)+\varepsilon_2 C_{24}.
\end{align}
The functions $\varepsilon_1$ and $\varepsilon_2$ are given as,  
\begin{align}
& \varepsilon_1=a\alpha T^{a}M^{b}[aT^{-1}K_{n}+K^{\prime}_{n}]\nonumber\\
    & ~~~~~+\alpha(a+1)T^a M^bK_n^{\prime}+\alpha T^{a+1}M^b K^{\prime\prime}_n,
\end{align}    

\begin{align}    
    & \varepsilon_2=\alpha T^a M^b (a K_n+ T K_n^{\prime}) .
\end{align}
Therefore the hydrodynamic equation governing the proper time evolution of temperature can be recast as, 
\begin{align}
    & \frac{d\varepsilon}{d\tau}+ \frac{\varepsilon+p}{\tau}-\frac{s_{0}}{\tau^2}\bigg(\frac{2}{3}\frac{\eta_s}{s_{0}}+\frac{\zeta}{s_0}\bigg)=0\nonumber\\
    \implies &     A(\tau) \frac{d^2T}{d\tau^2}+B(\tau)\bigg(\frac{dT}{d\tau}\bigg)^2+D(\tau)\frac{dT}{d\tau}+E(\tau)=0.
\end{align}
In the above equation, 
\begin{align}
    & A(\tau)= \varepsilon_2~C_{22},\nonumber\\
    & B(\tau)= (\varepsilon_2C_{21}+\varepsilon_1 C_{11}),\nonumber\\
    & D(\tau)= \bigg(\frac{d\varepsilon_{0}}{dT}+C_{12}\varepsilon_1+C_{23}\varepsilon_2\bigg)+\frac{\varepsilon_2}{\tau}C_{11}\nonumber\\
    & \quad~~~~~~~~~~~~~~~~~ +\frac{\alpha}{\tau}T^a M^{3-a}K_n~C_{11},\nonumber\\
    & E(\tau)= \varepsilon_2 C_{24}+\frac{\varepsilon_0+p_0}{\tau}+\frac{\varepsilon_2~C_{12}}{\tau}\nonumber\\
    &\quad ~~~~+\frac{\alpha}{\tau}T^aM^{3-a}K_n C_{12}-\frac{s_{0}}{\tau^2}\bigg(\frac{2}{3}\frac{\eta_s}{s_{0}}+\frac{\zeta}{s_0}\bigg).
\end{align}

\bibliography{ref.bib}{}
\bibliographystyle{utphys}

\end{document}